\documentclass{article}

\usepackage{PRIMEarxiv}

\usepackage[utf8]{inputenc} 
\usepackage[T1]{fontenc}    
\usepackage{hyperref}       
\usepackage{url}            
\usepackage{booktabs}       
\usepackage{amsfonts}       
\usepackage{nicefrac}       
\usepackage{microtype}      
\usepackage{lipsum}
\usepackage{fancyhdr}       
\usepackage{graphicx}       
\graphicspath{{media/}}     

\usepackage{caption}
\usepackage{subcaption}
\usepackage{algorithm}
\usepackage{algpseudocode}

\makeatletter
\renewcommand{\ALG@name}{Model}
\makeatother

\title{dFlow: A Domain Specific Language for the Rapid Development of open-source Virtual Assistants
}

\author{
  Nikolaos Malamas \\
  School of Electrical and Computer Engineering \\
  AUTH, Thessaloniki, Greece \\
\tt{nmalamas@ece.auth.gr} \\[3pt]
  Gnomon Informatics S.A. \\
  Thessaloniki, Greece \\  
   \And
  Konstantinos Panayiotou \\
  School of Electrical and Computer Engineering \\
  AUTH, Thessaloniki, Greece \\
  \tt{klpanagi@ece.auth.gr} \\
  \And
    Andreas L. Symeonidis \\
  School of Electrical and Computer Engineering \\
  AUTH, Thessaloniki, Greece \\
  \tt{symeonid@ece.auth.gr} \\[3pt]
  Cyclopt P.C. \\
  Thessaloniki, Greece \\  
}

\begin{document}
\maketitle

\begin{abstract}
An increasing number of models and frameworks for Virtual Assistant (VA) development exist nowadays, following the progress in the Natural Language Processing (NLP) and Natural Language Understanding (NLU) fields. Regardless of their performance, popularity, and ease of use, these frameworks require at least basic expertise in NLP and software engineering, even for simple and repetitive processes, limiting their use only to the domain and programming experts. However, since the current state of practice of VA development is a straightforward process, Model-Driven Engineering approaches can be utilized to achieve automation and rapid development in a more convenient manner. To this end, we present \textit{dFlow}, a textual Domain-Specific Language (DSL) that offers a simplified, reusable, and framework-agnostic language for creating task-specific VAs in a low-code manner. We describe a system-agnostic VA meta-model, the developed grammar, and all essential processes for developing and deploying smart VAs. For further convenience, we create a cloud-native architecture and expose it through the Discord platform. We conducted a large-scale empirical evaluation with more than 200 junior software developers and collected positive feedback, indicating that dFlow can accelerate the entire VA development process, while also enabling citizen and software developers with minimum experience to participate.
\end{abstract}


\section{Introduction}\label{sec1}

Virtual Assistants have been increasingly used in several well-defined and repetitive cases, customer support, FAQ systems, and smart home environments, among others. While systems like Apple Siri and Amazon Alexa have been around for several years, the recent advances in NLP/NLU techniques and tools, such as ChatGPT\footnote{https://openai.com/blog/chatgpt/} and LaMDA \cite{lambda}, have generated high hopes for the domain, even though they are not yet production-ready models. At the same time, we witness a steady shift and integration of NL-based interfaces from most of the conventional web- and mobile-based systems \cite{planas2021} and a plethora of conversational frameworks have been developed, both commercial and open-source. 

Despite the abundance of choices as presented in Harms et al. \cite{harms2019}, most of the frameworks tend to come with time-consuming, repetitive processes and little automation. They demand experienced developers with proper domain knowledge as these frameworks can be very complex. Even commercial systems that may offer more automation, lead to vendor lock-in, which is not desired in most cases.

Model-Driven Engineering (MDE) is a software development methodology that aspires to simplify and automate the production of such "complex" systems and software applications. First, the \textit{meta-model}, an abstract representation of the system, is defined. Then, a Domain-Specific Language (DSL) is created that specifies the meta-model's concepts and rules in its grammar. Finally, a generator component is implemented that can generate the intended software given a \textit{model}. We witness such low-code and no-code systems in several domains, in robotics \cite{r4aarchitecture,mde_robotics}, cyber-physical systems \cite{mde_cps,dsl_cps_pradhan}, and the Internet-of-Things \cite{mde_iot}, among others. In addition, DSLs can be grouped into textual and graphical, depending on their interface. While the latter might be considered easier for non-experts like citizen developers, i.e., people with basic programming skills and limited domain knowledge, the former is preferred when it comes to such end users as it is considered simpler and more practical \cite{xatkit}.

Even though there is work in other domains, using MDE approaches in the area of VAs is relatively new. We see only a handful of studies that thoroughly model Virtual Assistants and use MDE processes for rapid VA development, despite the fact that low-code and no-code methods are going to be of great importance in the coming years \cite{cabot20} and will enable access to a wider range of end-users and citizen developers. This way, tech companies will acquire an increasing number of content creators apart from experienced software developers. In addition and to our knowledge, no studies employ entirely open-source tools, thus the threat of vendor lock-in is quite eminent, as stated above.

To this end, we present \textit{dFlow}, a textual DSL that offers a simplified, reusable, and system-agnostic process for creating task-specific VAs in a low-code manner. We demonstrate our VA meta-model on which dFlow is based. We also develop and present a cloud architecture and its processes deployed via the Discord platform, with which users regardless of their development skills can implement, test, and deploy assistants. The generated assistants employ the Rasa framework as it is an open-source and well-performing system \cite{malamas-healthcare}. 

The paper is structured as follows. Section \ref{sec:related_work} discusses the state-of-the-art conversational models and frameworks, followed by studies that employed DSLs to create conversational systems. In section \ref{sec:3}, the grammar of dFlow with its concepts is presented. In addition, the cloud-native architecture is depicted with the implemented processes and automations. Section \ref{sec:4} lists the research questions of this work, presents a large workshop conducted with more than 200 young programmers, and discusses the collected evaluation and feedback. Finally, Section \ref{sec:5} probes into conclusions and future work.

\section{Related Work}\label{sec:related_work}

\subsection{Conversational Models and Frameworks}

The release of ChatGPT is undoubtedly a cornerstone for the Conversational AI and NLP domain. It is a conversational model trained using a combination of supervised learning and reinforcement learning techniques from human feedback (RLHF) \cite{rlhf}, and is built upon GPT-3 \cite{gpt3}, a large autoregressive Language Model (LM) that can generate text from a given prompt in a few-shot manner. The produced text, however, is not preferred for dialogue as studies showed \cite{instructgpt}; for this reason, InstructGPT models are developed, which are fine-tuned GPT-3 models that can follow user instructions and generate more fluent and coherent results. In addition, GPT-4 \cite{openai2023gpt4}, the latest GPT-based model, achieved human-level performance on several professional and academic benchmarks. Of similar impact to the domain is the development of LaMDa \cite{lambda}. LaMDa is also a conversational LM that can be grounded in external knowledge sources and produce more factual results. 

Although impressive and efficient, one important limitation of these models is the fact that they cannot access external knowledge sources (e.g., the web) in real time to retrieve data and produce up-to-date results. There are several studies that identify this limitation and offer solutions. For example, the BlenderBot 2.0 \cite{blenderbot2} conversational model learns to generate appropriate Internet queries from user expressions and uses the results to enrich its responses. Retro \cite{retro} on the other hand, is a BERT-based Retriever-Enhanced model that can access an external database and retrieve relevant information to user queries and answer their questions. Finally, in another study, GPT-3 was a fine-tuned model to create WebGPT \cite{webgpt}. This model is trained to search the Internet, follow site links, and scroll the accessed web pages in order to generate the best answer to each question while citing its sources.

However promising their results are, these generative models are not yet production-ready systems for a variety of reasons. First, they tend to hallucinate and generate misinformation that seems rather factual \cite{Metzler_2021}. We have witnessed Galactica \cite{galactica}, a very large LM that can generate seemingly true scientific papers, raise important ethical questions regarding misinformation, and was taken down almost immediately. Next, as stated earlier, data grounding can be accomplished with web searching or database querying, but their results are not always predictable, while very thorough prompt engineering is needed to achieve optimal responses. Another very important aspect that cannot yet be tackled by those Language Models is the interaction with the human world in a more complex manner than web searches like smart systems and environments. In these cases, Virtual Assistants should be able to manage users' agendas, set activities, offer recommendations, or access other applications and devices. The current state of practice in these situations is to employ frameworks and determine the closed domain in which the assistant will operate. Ergo, the assistants will have predictable behavior by only performing the above-mentioned actions.

There is a variety of frameworks to create conversational task-based systems that are considered the current state of practice. Regarding NLU, most frameworks achieve quite similar performance as demonstrated in various studies \cite{braun2017,Liu2019BenchmarkingNL,Qi2020BenchmarkingID}, hence the final selection is primarily determined considering the functionality offered by each framework and the intended use case. For example, in cases where there is a need for independence from all deployment processes, automation, or efficient scalability, commercial frameworks such as Google Dialogflow, Amazon Lex, and IBM Watson, are well-featured choices. This way developers can focus only on creating and optimizing the assistant. In addition, these frameworks usually offer both programming libraries and graphical interfaces. Furthermore, they are the only choice when the produced assistant is going to be embedded in proprietary devices like Amazon Alexa or Google Home, leading to more efficient integration. Despite their convenience, however, these devices have raised important questions regarding privacy and security, while these frameworks do not promise careful data usage and also lead to vendor lock-in \cite{Lei2018,malamas2021-rasa}. 

It is common knowledge that the most efficient way to enable users to create chatbots and Virtual Assistants is graphical DSLs (Domain-Specific Languages), through platforms-as-a-service tools. Typical cases of such platforms include: fabble\footnote{https://fabble.io/}, Flow XO \footnote{https://flowxo.com/}, moveo\footnote{https://moveo.ai/}, botpress\footnote{https://botpress.com/}, and voiceflow\footnote{https://www.voiceflow.com/}, to name a few. These platforms, however, usually offer limited functionality, access, or amount of traffic in their free plans and force developers to follow paid plans.

To this end, open-source Virtual Assistant frameworks, such as Rasa \cite{rasa}, ChatterBot\footnote{https://github.com/gunthercox/ChatterBot}, and wit.ai\footnote{https://wit.ai/} are a good alternative for avoiding lock-ins and expanding functionality. Particularly, they are fully transparent and self-hosted systems that can handle and store user data in a secure manner and can even operate on the edge \cite{malamas2021-rasa}. In addition, their given customization can benefit both the developers who are able to produce more flexible systems and the users who can use personalized assistants \cite{malamas2021-rasa}. Nevertheless, these frameworks usually require high programming skills and NLP expertise, which lead to time- and labor-intensive development processes, even for simple repetitive scenarios. Thus, the need for automating processes and allowing citizen developers to build their own scenarios is imminent.

\subsection{Domain-Specific Languages for Virtual Assistants}

In order to tackle the problem of framework-specific knowledge, one solution is to develop an abstract, system-agnostic meta-model of the Virtual Assistant and develop the associated DSL to declare and auto-generate assistants. Domain-specific languages have been around for many years and they are used to solve problems in various scientific and engineering fields, such as the domains of smart home \cite{desolda2017empowering} and smart grids \cite{adolf2012smartscript}. As far as DSL development is concerned, MDE \cite{kent2002model} (model-driven engineering) is a promising paradigm having the potential to overcome heterogeneity issues (i.e., platforms, languages and communication protocols, control and monitoring mechanisms). Using layered abstractions, MDE automates software engineering processes \cite{kent2002model} and enables partial or even complete generation of systems via Model-to-Model and Model-to-Text transformations.

DSLs, in specific, are languages tailored to a specific application domain \cite{mernik2005and} and therefore have the potential to reduce development complexity by raising the abstraction level towards an application domain. Nowadays, DSL engineering approaches are utilized for the development of low-code \cite{sahay2020supporting} and no-code \cite{cypher2010no} solutions (e.g., frameworks and platforms) and enable access to a wide range of users and application domains. According to the application domain and the target group of users, different notations (textual, graphical, tabular) are used. DSL solutions provide an abstraction layer and user-oriented semantics to describe various aspects of the application domain(s) \cite{voelter2013dsl}. Furthermore, they implement verification so that the user can ensure that certain properties of the understudy domain/system/application hold true, via a set of meta-models that describe the domain. Model transformation and model interpretation processes are often delivered to enable the automated generation and deployment of software artifacts, based on input models \cite{brambilla2017model}.

One of the first complete VA DSLs is Xatkit \cite{xatkit,planas2021} which is an open-source textual framework for VA development designed in a three-package manner. The first two packages model the understanding and the response schema of the assistant, while the third is a runtime package for channel and platform-agnostic deployment. Xatkit has been used in several studies. For example, Xatkit was employed to automatically generate conversational interfaces on Open Data resources \cite{ed-douibi}. Particularly, it first retrieves and preprocesses the Open Data to a bot-friendly format, and then it generates a complete assistant that can select the most appropriate data source, process the results, and present them to the user. Similarly, Xatkit and Dialogflow were used to create systems to query multi-dimensional data through an NL interface \cite{Franciscatto}.

Furthermore, in another study, a VA to conversationally create other VAs was designed \cite{perez19}, in a so-called Conversation-as-a-Platform schema, using two separate meta-models. The first one is a meta-model of a VA implemented with Dialogflow, while the second one represents the NL syntax that the first assistant will understand and use to generate new interactive systems. Moreover, wcs-ocaml \cite{reactiveml} is a library for creating multi-tier assistants based on IBM Watson in a very programming fashion. As this method is entirely tailor-made for Watson, the problem of vendor lock-in still strongly exists. In addition, CONGRA \cite{congra} (ChatbOt modelliNG lanGuAge) is a platform-agnostic DSL that models VAs. It offers a web UI for more convenience \cite{congra_web} and it can also recommend the most suitable VA tool to generate each model according to a suitability score determined from a list of questions. For demonstration purposes, generator modules were developed for Dialogflow and Rasa using Xtext\footnote{https://www.eclipse.org/Xtext/}. While most presented meta-models are well-defined and can be replicated and used by the community, most of their existing implementations employ commercial VA frameworks, such as Google Dialogflow and IBM Watson, which as discussed earlier limit users from fully exploiting the benefits of open-source technologies. Consequently, dFlow was developed, which is a textual DSL for developing and deploying task-specific Virtual Assistants using entirely open-source tools following a low-code approach. It is also worth noting that most of the aforementioned systems are usually validated by a small number of participants, around 20, and in some cases none. On the contrary, dFlow was evaluated from 219 participants, which can be considered a significantly large study.

\section{Declarative Approach for Building Virtual Assistants}
\label{sec:3}

In this section, we present the concepts of our meta-model and its grammar, the implemented processes for model validation and code generation, and finally, the cloud architecture for VA development and testing. TextX\footnote{https://textx.github.io/textX/3.0/} has been employed for meta-model definition and grammar development, while the VA models are transformed into ready-to-deploy Rasa models.

\subsection{Modeling with dFlow}

\begin{figure*}[!h]
\centering
\includegraphics[width=0.8 \linewidth]{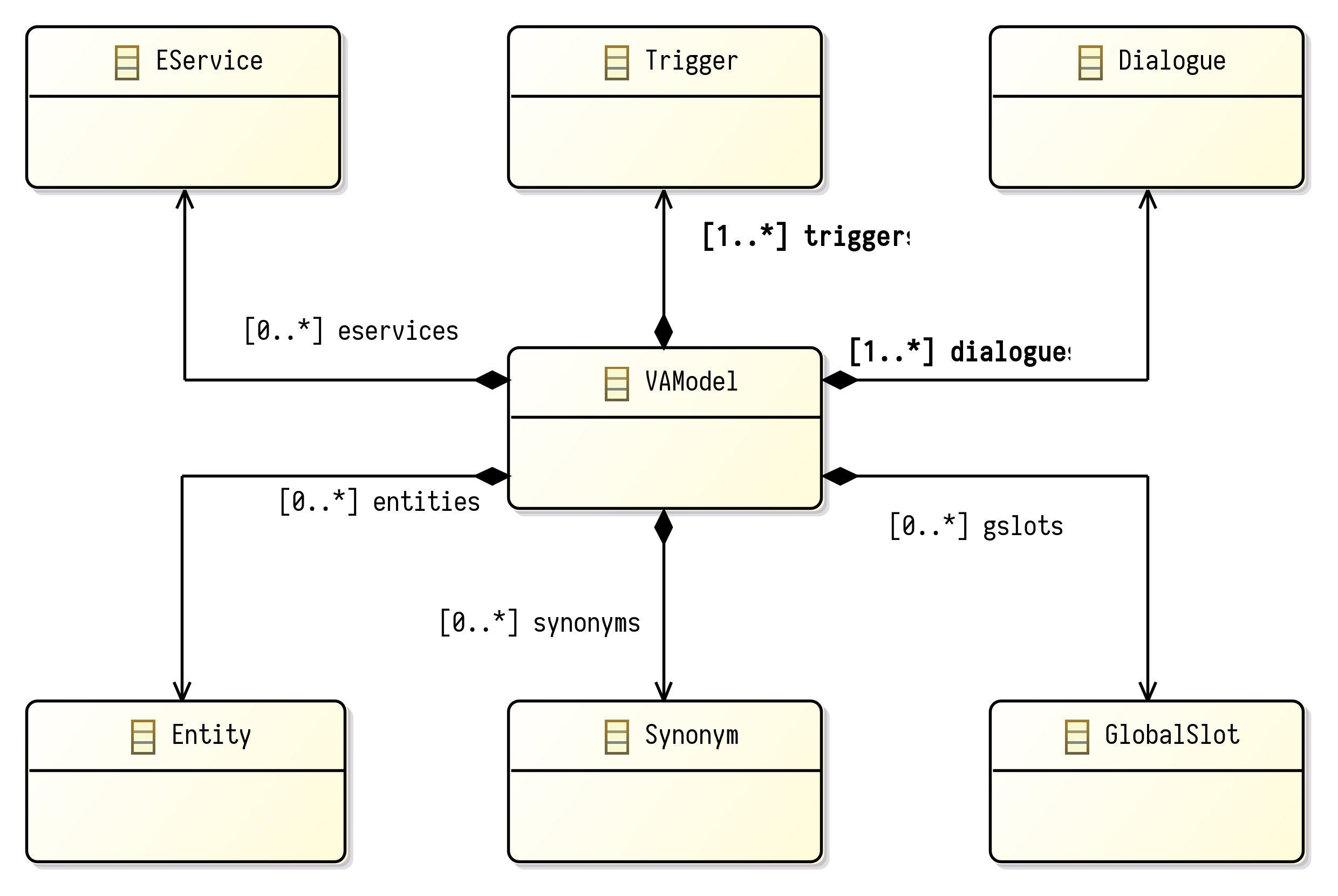}
\caption{dFlow Root Meta-model}
\label{fig:root_mm}
\end{figure*}

A task-oriented Virtual Assistant incorporates (among others) the following two components: the Natural Language Understanding (NLU) component, which is responsible for processing user utterances and interpreting the user's goals or intents, and the Natural Language Generation (NLG) part which is responsible for creating the most appropriate responses and actions. The root meta-model of dFlow consists of six (6) concepts as depicted in Figure \ref{fig:root_mm}: Entity, Synonym, Trigger, EService, Global Slot, and Dialogues. The first three concepts capture the NLU part of the assistant, the next two define reusable features in general scope, and the last one describes the dialogue flows and the assistant responses, hence the entire NLG component. A dFlow model incorporates these concepts at the root scope and can be utilized to define the interactive part and include bot responses to predefined conditions (e.g., an internal intent is triggered). All these are discussed next. 

\subsubsection{Entities \& Synonyms}

\begin{figure*}[!h]
\centering
\includegraphics[width=0.8 \linewidth]{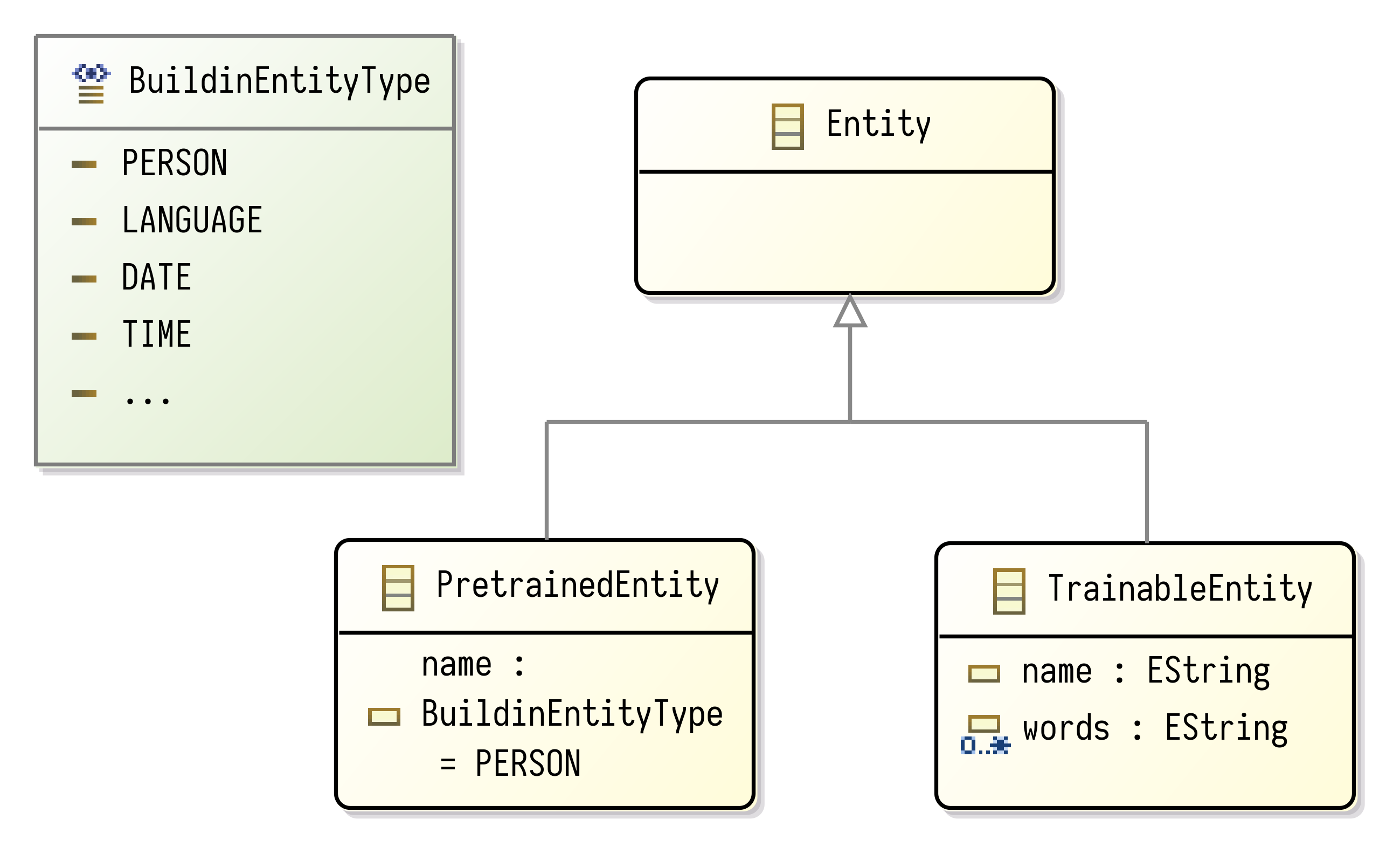}
\caption{dFlow Entity Meta-model}
\label{fig:entity_mm}
\end{figure*}

\textit{Entities} are structured pieces of information inside a user message that can be extracted and used by the assistant. They can be real-world objects or meanings, such as a person, a location, an organization, or a product. DFlow can employ already \textit{Pre-trained} Named Entity Recognition (NER) models that can efficiently extract those types of entities without further training. Quite frequently Virtual Assistants need to detect use-case-specific information that is not supported by the Pre-trained NER models, such as types of food or fruits. In this case, \textit{Trainable Entities} can be specified and trained during deployment given a set of entity examples. The Entity meta-model is presented in Figure \ref{fig:entity_mm}.

On the other hand, \textit{Synonyms} map words to a value other than the literal text extracted. They can be used when users refer to the same thing in multiple ways and no semantic difference exists between them.

\subsubsection{Triggers}

\begin{figure*}[!h]
\centering
\includegraphics[width=0.8 \linewidth]{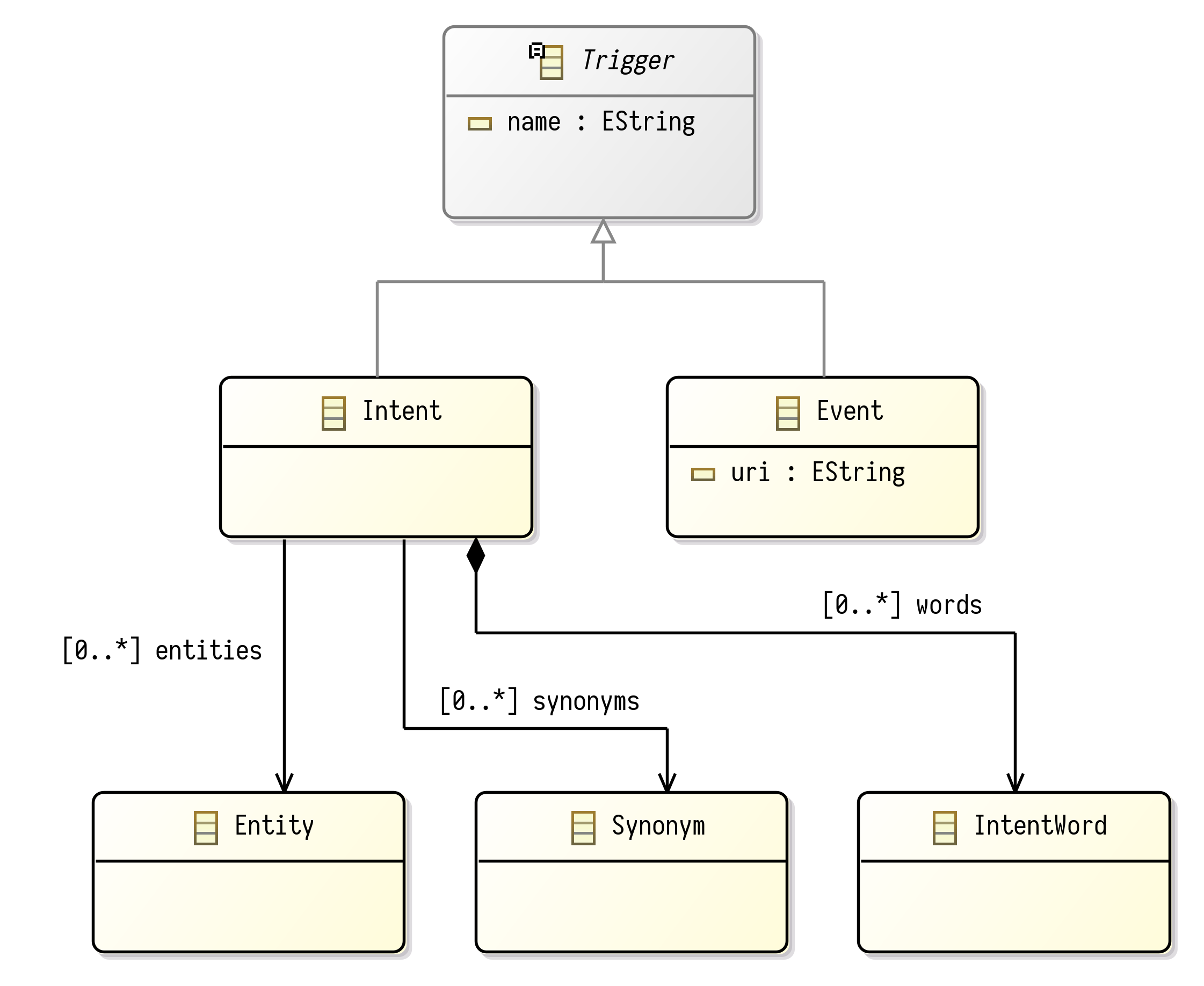}
\caption{dFlow Trigger Meta-model}
\label{fig:trigger_mm}
\end{figure*}

\textit{Triggers} represent the two ways a dialogue can be initialized: when a user states a particular expression or \textit{Intent}, or when an external \textit{Event} is triggered, such as a reminder or a notification. In more detail, in task-based dialogue systems, an Intent is a goal the user is trying to achieve or accomplish, such as retrieving specific information on the weather or setting a reminder. An \textit{Intent} needs a set of phrase examples, that are semantically similar to the expected user expressions. An important aspect of the examples is that they can consist of combinations of text, Pre-trained and Trainable Entities, as well as Synonyms, as shown in Figure \ref{fig:trigger_mm}. On the other hand, Events are system-initiated and do not need any phrase examples.

\subsubsection{EServices}
\begin{figure*}[!h]
\centering
\includegraphics[width=1 \linewidth]{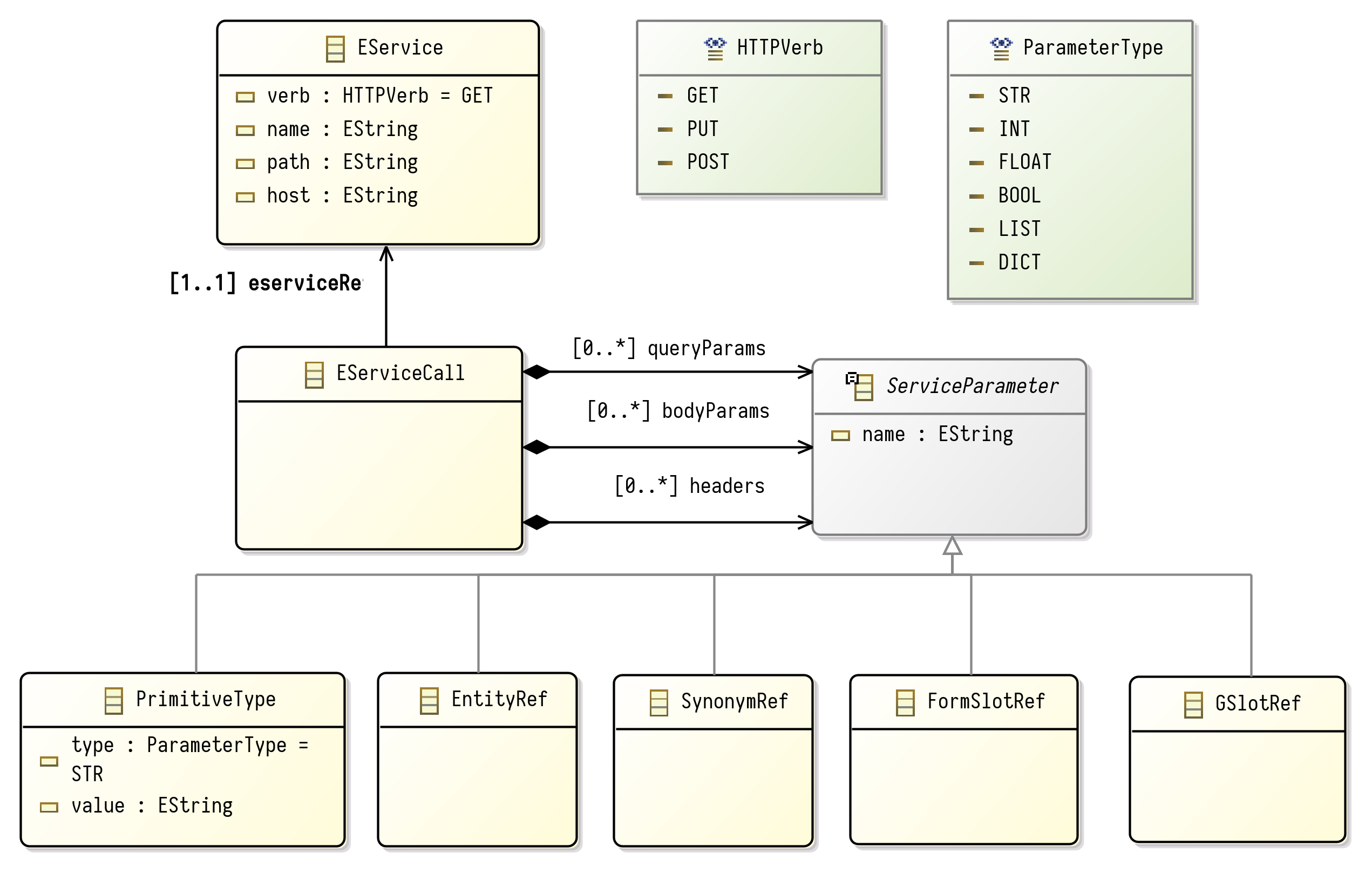}
\caption{dFlow EService Meta-model}
\label{fig:service_mm}
\end{figure*}

External \textit{services} are REST endpoints that can be used as part of the VA's responses. Their URL and HTTP method are defined globally as static attributes, while their parameters can be specified inside the dialogue section when called, in a more dynamic manner as depicted in Figure \ref{fig:service_mm}.

\subsubsection{Global Slots}

Slots are static information an assistant can access and use, a feature that can offer multi-turn conversations, memory, and personalization. DFlow introduces the \textit{GSlot} concept that can be used to define variables in the global scope so that they can be accessed by various Dialogues, Forms, and Actions, as will be discussed in the next paragraphs.

\subsubsection{Dialogues}

\begin{figure*}[!h]
\centering
\includegraphics[width=1 \linewidth]{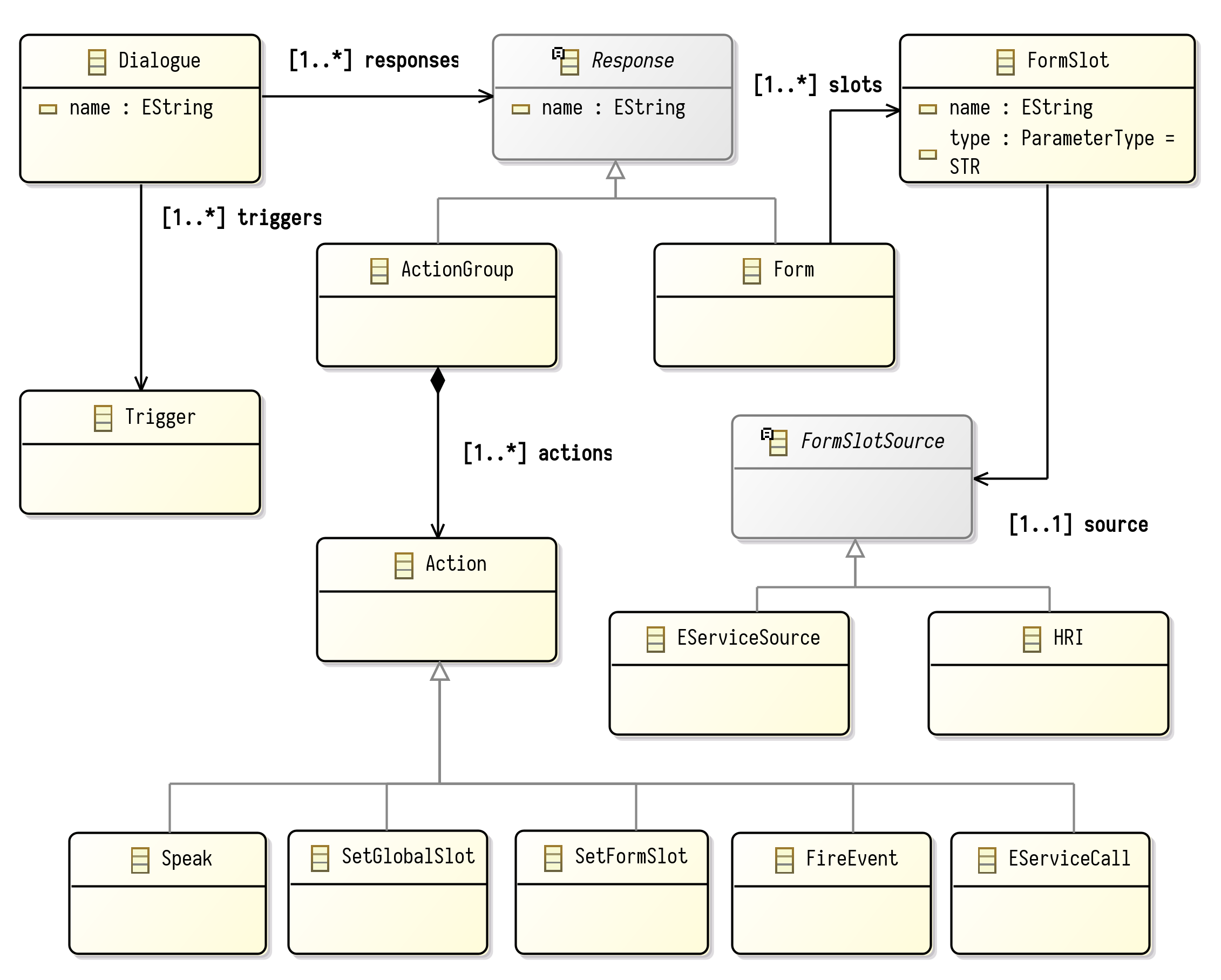}
\caption{dFlow Dialogue Meta-model}
\label{fig:dialogue_mm}
\end{figure*}

An important concept of the dFlow meta-model is the \textit{Dialogue}. Dialogues are conversational flows the assistant supports in the form of trigger and response pairs, where each response is a sequence of \textit{Forms} and \textit{ActionGroups} in a one-turn conversation manner. It is noted that each trigger initiates only one dialogue since more complex scenarios should also consider the conversation history, which is out of the scope of the current dFlow version as presented in Figure \ref{fig:dialogue_mm}.

\begin{figure*}[!h]
\centering
\includegraphics[width=0.7 \linewidth]{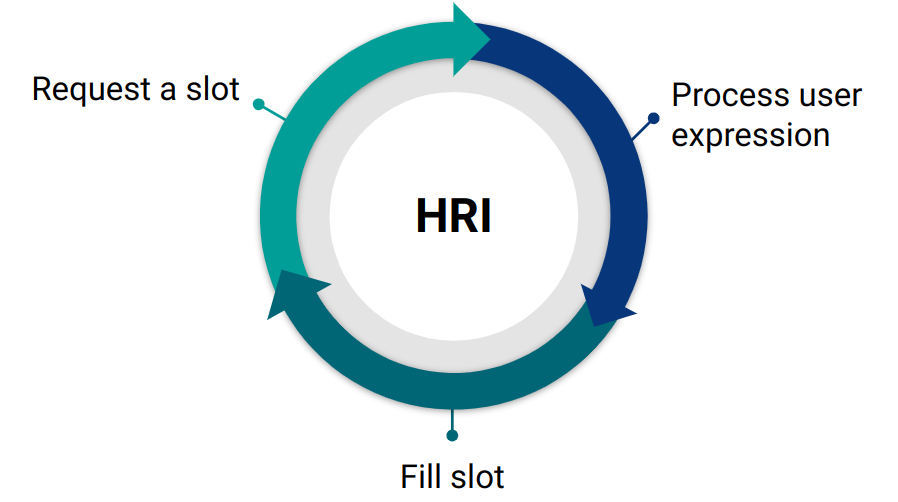}
\caption{Human-Robot Interaction Flow using the Slot concept}
\label{fig:hri}
\end{figure*}

Regarding the responses, a \textit{Form} is a conversational pattern to collect information and store it in form parameters, or \textit{form slots}, following business logic. Two interaction methods are currently supported by the dFlow DSL, Human-Robot Interaction or \textit{HRI} and \textit{External Services}. Information can be collected via a Human-Robot Interaction or \textit{HRI}, where the assistant sequentially collects the information from the user. It requests each slot using a specified text and extracts the data from the user expression as presented in Figure \ref{fig:hri}. The expression can be defined to contain the entire text, an extracted entity, or a specific value set mapped to a particular intent. The second choice is the \textit{EServiceSource} interaction, where the slot is filled with information received from an external service, that is previously defined as an \textit{EService} in the dFlow model.

Furthermore, an \textit{ActionGroup} is a set of \textit{Actions}. The dFlow language supports five different types of actions; the assistant can state a given phrase (\textit{SpeakAction}), fire a broker event (\textit{FireEventAction}), call a REST endpoint (\textit{RESTCallAction}), set a global slot (\textit{SetGSlot}) or form slot (\textit{SetFSlot}), with particular parameters. Actions can also use real-time environment parameters and functions grouped as \textit{UserProperties} and \textit{SystemProperties}. User properties are user information stored locally in the device that the assistant can use, such as the name, surname, age, email, phone, city, and address. System properties, on the other hand, are in-built system functions to get the current time, location, and a random integer or float. That way the assistant has access to those data being device-agnostic at the same time while offering more dynamic and personalized dialogues.

\subsection{Software Automation}

In the context of the current work a Model-to-Text (M2T) transformation has been developed that enables the automated generation of ready-to-deploy virtual assistants. The M2T takes a dFlow model as input and uses the Rasa framework \cite{rasa} to build the assistant and integrate dialogue flows defined by the user. The M2T transformation processes and maps each dFlow attribute to one or more corresponding Rasa attributes, as the Rasa file structure consists of several configuration files and Python scripts for static and dynamic behavior, respectively. For example, a simple \textit{hello-world} one-dialogue VA implemented on Rasa would need eight different files and a total of 52 lines of code while it would require 16 lines of dFlow as presented in Model \ref{alg:hello_world}.

\begin{algorithm}[htb]
\caption{Hello world}\label{alg:hello_world}
\begin{algorithmic}[1]
\State triggers
\State \quad  Intent greet
\State \quad \quad "hello",
\State \quad \quad "hey",
\State \quad \quad "good morning"
\State \quad end
\State end

\State dialogues
\State \quad  Dialogue dialogue\_1
\State \quad \quad on: greet
\State \quad \quad responses:
\State \quad \quad \quad ActionGroup resp
\State \quad \quad \quad \quad Speak('Hello friend')
\State \quad \quad \quad end
\State \quad end
\State end
\end{algorithmic}
\end{algorithm}

Furthermore, the dFlow DSL implements a model validator, that is used to validate dFlow instance models against various relational and logical rules defined by the meta-model (models must conform to the meta-model) and a reasoning process. Model validation can be executed both at development time, just like any other language syntax checker, and before performing the M2T. The validation process is considered important and minimizes errors beyond simple syntax issues since domain-specific knowledge is defined by the meta-model that dFlow models must conform to.

\subsection{Cloud-native Development and Continues Delivery}

\begin{figure*}[!h]
\centering
\includegraphics[width=0.6 \linewidth]{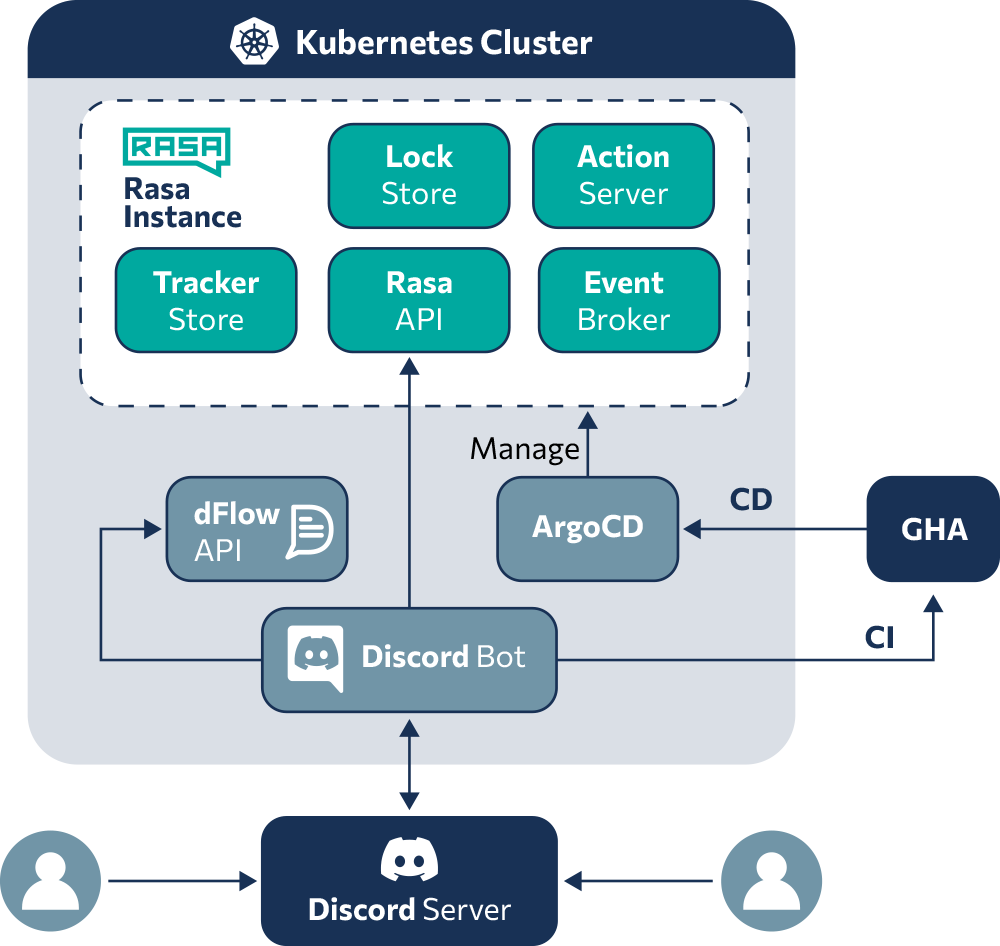}
\caption{Architecture of the automated build and deploy platform}
\label{fig:arch}
\end{figure*}

All systems have been transferred to the cloud for location-agnostic development and deployment of Virtual Assistants from citizen and software developers. This cloud-native approach comprises several components as depicted in Figure \ref{fig:arch}. The core ones are the REST API of dFlow, as well as a Rasa deployment that includes several services (action server, database, event broker, telemetry, Rasa API for remote interaction, lock store, and tracker store). Everything is deployed within the cluster and GitOps via Github Actions (GHA) is used to automate the build and deployment of Rasa instances within the cluster. Finally, a Discord bot that connects user interfaces to the aforementioned backend services has also been deployed in the cluster.

Basic dFlow operations (e.g., validation and code generation) have been RESTified to provide a number of endpoints in the form of an API based on the OpenAPI specification\footnote{https://spec.openapis.org/oas/v3.1.0}. A model can be validated against the domain meta-model and its constraints, generated to a ready-to-deploy VA via an M2T transformation, retrieved, stored, or updated while keeping the older models. The API also provides an endpoint to retrieve the last stored model of a user, and finally, an endpoint that enables merging the last submitted models, from all registered users, into a single model that includes all dialogue flows defined within the individual dFlow models. Model merging can be used in collaborative development schemes for multi-user VAs. Furthermore, an SQL database has been integrated into the API backend for storage and retrieval of instance models submitted by registered users, as well as for storing user accounts and for performing authentication and authorization for the endpoints. Table \ref{tab:dflow-api} documents the list of the essential endpoints that compose the dFlow REST API. Notice that user-related endpoints used for authentication and authorization (such as for account creation, user login, etc.) are not documented, since these are considered out of the scope of the current manuscript.

\begin{table}[]
\centering
\begin{tabular}{lll} 
\toprule
\textbf{Endpoint} & \textbf{Verb} & \textbf{Description} \\ \midrule
/model & POST & Store Model in DB                      \\
/model/\{model\_id\} & GET, PUT, & Retrieve/Update/Remove    \\
 & DELETE & model from DB \\
/model/validation & POST & Perform validation on       \\
 & & input model \\
/model/codegen & POST & Perform code generation  \\
 & & on input model \\
/model/merge & GET & Retrieve merged model \\
/user/\{username\}/model/latest & GET & Retrieve latest model  \\
 & & for user \\
\bottomrule
\end{tabular}
\caption{DFlow model-related API endpoints}
\label{tab:dflow-api}
\end{table}

Next, a Rasa VA stack has been deployed within a Kubernetes\footnote{https://kubernetes.io/} cluster. Rasa provides several in-built REST endpoints, three of which are utilized to perform operations and interact with VA instances. Specifically, a) the \textit{Train} endpoint is used to receive a complete Rasa model and perform the internal training process, b) the model activation endpoint to automatically deploy the latest trained model, and c) the \textit{Dialogue} endpoint to chat with a running VA instance. Rasa also employs an optional custom Action Server, a Docker image containing Python scripts responsible for dynamic assistant responses. As all dFlow models generate such types of responses, it is mandatory for the deployment of our models. Hence, the appropriate automated continuous integration and continuous delivery processes have been implemented using Github Actions\footnote{https://docs.github.com/en/actions} for automated build and upload of the software artifacts and ArgoCD\footnote{https://argoproj.github.io/}, a declarative continuous delivery tool for cloud-based applications, for automatically uploading the deployments that are triggered from successful execution of the GHA beforehand.

Finally, the most important component is the interface that allows users to interact with dFlow, validate, generate, and deploy VAs. Discord has been selected as a simple messaging platform in which both development and testing can occur. The DiscordPy library\footnote{https://discordpy.readthedocs.io/en/stable/} has been employed for the implementation of the dFlow discord bot, which supports several keyword-triggered commands for accessing the aforementioned REST API of dFlow. In case no command is detected, it is assumed that the input message is an actual conversation and it is routed to the activated Rasa model for direct interaction with the previously deployed VA.

\subsection{Development and Deployment Process Demonstration}

\begin{figure*}[!h]
\centering
\includegraphics[width=0.9 \linewidth]{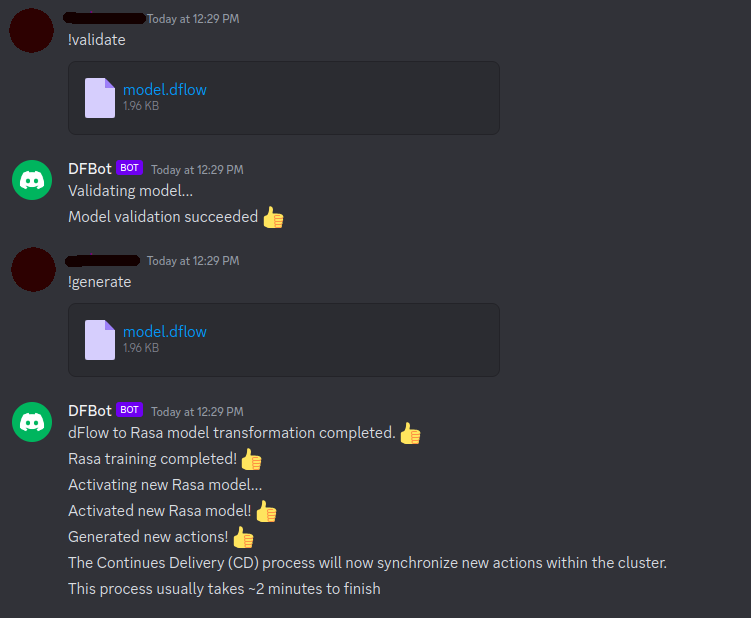}
\caption{Example of the automated build and deploy platform}
\label{fig:discord-example}
\end{figure*}

\begin{figure*}[!h]
\centering
\includegraphics[width=0.5 \linewidth]{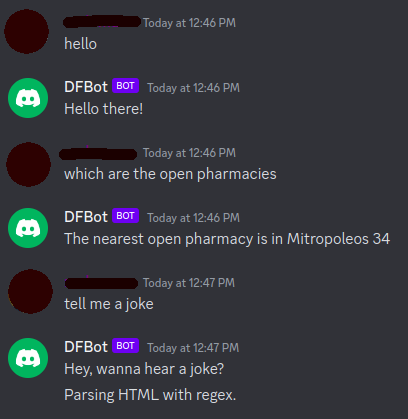}
\caption{Example conversation with the deployed assistant}
\label{fig:discord-conversation}
\end{figure*}

The entire cloud-native architecture can be better demonstrated with the following example. A Virtual Assistant has been designed that supports three scenarios: greeting the user, finding the nearest open pharmacies, and telling jokes. These scenarios require three intents (Model \ref{alg:complex_assistant_triggers}), three external services (Model \ref{alg:complex_assistant_eservices}), and three dialogues (Model \ref{alg:complex_assistant_dialogues}). The dFlow model, which is the concatenation of the three presented models, is uploaded to the Discord platform for validation to first check its syntax. Then, as the validation is successful, it is sent for generation and deployment. This interaction is also depicted in Figure \ref{fig:discord-example}. The dFlow model consists of one file of 72 lines of code, while the generated Rasa model consists of different 8 files with a total of 357 lines of code. After a few minutes, the Discord bot notifies us that the deployment of the generated model to the Rasa cluster is completed, and hence, we can interact with it and conduct the conversation presented in Figure \ref{fig:discord-conversation}.

\begin{algorithm}[htb]
\caption{Demonstration assistant Triggers}\label{alg:complex_assistant_triggers}
\begin{algorithmic}[1]

\State triggers
\State \quad Intent greet
\State \quad \quad "hey",
\State \quad \quad "hello there",
\State \quad \quad "good morning",
\State \quad \quad "good afternoon",
\State \quad \quad "what's up"
\State \quad end

\State \quad Intent find\_pharmacy
\State \quad "which pharmacy is open today",
\State \quad \quad "I need to go to the pharmacy right now",
\State \quad \quad "I have to buy new medicine",
\State \quad \quad "open pharmacies"
\State \quad end

\State \quad Intent tell\_joke
\State \quad \quad "tell me a joke",
\State \quad \quad "do you know any good jokes",
\State \quad \quad "time for a laugh",
\State \quad \quad "make me laugh"
\State \quad end
\State end

\end{algorithmic}
\end{algorithm}

\begin{algorithm}[htb]
\caption{Demonstration assistant Eservices}\label{alg:complex_assistant_eservices}
\begin{algorithmic}[1]

\State eservices
\State \quad EServiceHTTP coords\_svc
\State \quad \quad verb: GET
\State \quad \quad host: 'http://services.issel.ee.auth.gr'
\State \quad \quad path: '/geospatial/get\_coords'
\State \quad end

\State \quad EServiceHTTP pharmacy\_svc
\State \quad \quad verb: GET
\State \quad \quad host: 'http://services.issel.ee.auth.gr'
\State \quad \quad path: '/medical/pharmacies\_nearest'
\State \quad end

\State \quad EServiceHTTP jokes\_svc
\State \quad \quad verb: GET
\State \quad \quad host: 'http://services.issel.ee.auth.gr'
\State \quad \quad path: '/quotes/get\_joke'
\State \quad end
\State end

\end{algorithmic}
\end{algorithm}

\begin{algorithm}[htb]
\caption{Demonstration assistant Dialogues}\label{alg:complex_assistant_dialogues}
\begin{algorithmic}[1]

\State dialogues

\State \quad Dialogue greet\_dialogue
\State \quad \quad on: greet
\State \quad \quad responses:
\State \quad \quad \quad ActionGroup greet\_back
\State \quad \quad \quad \quad Speak('Hello there!')
\State \quad \quad \quad end
\State \quad end

\State \quad Dialogue pharmacy\_dialogue
\State \quad \quad on: find\_pharmacy
\State \quad \quad responses:
\State \quad \quad  \quad Form form1
\State \quad \quad \quad  \quad lan: str = coords\_svc(query=[place=USER:CITY], header=[access\_token="TOKEN"],)[lan]
\State \quad \quad \quad  \quad lon: str = coords\_svc(query=[place=USER:CITY], header=[access\_token="TOKEN"],)[lon]
\State \quad \quad \quad  \quad pharmacy\_slot: str = pharmacy\_svc(query=[latitude=form1.lan, longtitude=form1.lon], 
\State \quad \quad \quad \quad \quad header=[access\_token="TOKEN"],)[data.address]
\State \quad \quad \quad end,
\State \quad \quad \quad ActionGroup answer\_back
\State \quad \quad \quad \quad Speak('The nearest open pharmacy is in ' form1.pharmacy\_slot)
\State \quad \quad \quad end
\State \quad end

\State \quad Dialogue joke\_dialogue
\State \quad \quad on: tell\_joke
\State \quad \quad responses:
\State \quad \quad \quad Form form2
\State \quad \quad \quad \quad question: str = jokes\_svc(query=[language="English"], header=[access\_token="TOKEN"],)[question]
\State \quad \quad \quad \quad answer: int = jokes\_svc(query=[language="English"], header=[access\_token="TOKEN"],)[answer]
\State \quad \quad \quad end,
\State \quad \quad \quad ActionGroup say\_joke
\State \quad \quad \quad \quad Speak(form2.question)
\State \quad \quad \quad \quad Speak(form2.answer)
\State \quad \quad \quad end
\State \quad end
\State end

\end{algorithmic}
\end{algorithm}

\section{Empirical Evaluation}
\label{sec:4}

As already discussed, dFlow aims to automate the process of Virtual Assistant development and to include non-experts, i.e., citizen developers. We evaluate our approach against three research questions. 

\begin{itemize}
    \itemsep 0em
    \item \textbf{R.Q.1} Can we accelerate the development of Virtual Assistants?
    \item \textbf{R.Q.2} Does our approach need minimum domain knowledge and is it framework- and hardware-agnostic?
    \item \textbf{R.Q.3} Is the cloud architecture a convenient interface for development and deployment?
\end{itemize}

\subsection{Evaluation results}

We conducted a large-scale 2-section workshop, of around 2 hours each, to present and evaluate dFlow, where 219 university students participated. Some were familiar with programming, a few of them with Python and DSLs, but most of them had limited programming knowledge and no knowledge of NLP or Virtual Assistants, an overall background profile similar to the average citizen developer. In the first part, a detailed presentation of the basic concepts of NLP, DSLs, and Virtual Assistants, as well as the grammar of dFlow was given. In the second part, 3 example use case assignments were given to the participants to experiment and create VAs via Discord. The three given scenarios were a simple \textit{hello-world} assistant, an assistant for filling a user profile, and a weather forecast assistant. A dFlow model of the last and more complex scenario is presented in Model \ref{alg:weather_assistant} in more detail.

\begin{algorithm}[htb]
\caption{Weather assistant}\label{alg:weather_assistant}
\begin{algorithmic}[1]

\State triggers
\State \quad Intent ask\_weather
\State \quad \quad "I want to tell me the weather",
\State \quad \quad "Tell me the weather please",
\State \quad \quad "I want to tell me the weather for" PE:GPE['Thessaloniki', 'Athens'],
\State \quad \quad "Tell me the weather please for" PE:DATE['tomorrow', 'today'],
\State \quad \quad "Tell me the weather" PE:DATE['tomorrow', 'today'] "for"  PE:GPE['Thessaloniki', 'Athens']
\State \quad end
\State end

\State eservices
\State \quad EServiceHTTP weather\_svc
\State \quad \quad verb: GET
\State \quad \quad host: 'http://services.issel.ee.auth.gr'
\State \quad \quad path: '/general\_information/weather\_openweather'
\State \quad end
\State end

\State dialogues
\State \quad Dialogue weather\_dialogue
\State \quad \quad on: ask\_weather
\State \quad \quad responses:
\State \quad \quad \quad Form form1
\State \quad \quad \quad \quad city\_slot: str = HRI('For which city?', [PE:GPE])
\State \quad \quad \quad \quad answer: str = weather\_svc(query=[city=form1.city\_slot, language="English"], 
\State \quad \quad \quad \quad \quad header=[access\_token=TOKEN],)[description]
\State \quad \quad \quad end,
\State \quad \quad \quad ActionGroup answer\_back
\State \quad \quad \quad \quad Speak('The weather for' form1.city\_slot ' is ' form1.answer)
\State \quad \quad \quad end
\State \quad end
\State end

\end{algorithmic}
\end{algorithm}

After concluding the workshop, a questionnaire was handed to each participant to report their experience and give feedback on the dFlow and the developed processes. Overall, most participants were satisfied with the simplicity of the language, the existing documentation, and the convenient cloud-based interface. Several people who were not so familiar with programming found the documentation very extensive and in some cases chaotic. In addition, most participants described the validation process as very important and well-performing. However, dFlow does not currently support optimal error handling, thus debugging proved hard when complex errors were made. Additionally, the process for deploying the VAs is currently time-consuming (requires a few minutes) as the existing technologies are not yet able to process and deploy VA models instantly, thus users could not quickly test their assistants end-to-end. Finally, there were a few recommendations for further extending the dFlow functionalities, such as conditional logic and better manipulation of the external services.

\begin{figure}[!htb]\captionsetup[subfigure]{font=footnotesize}
    
\begin{subfigure}{.49\textwidth}
  \centering
  \includegraphics[width=\linewidth]{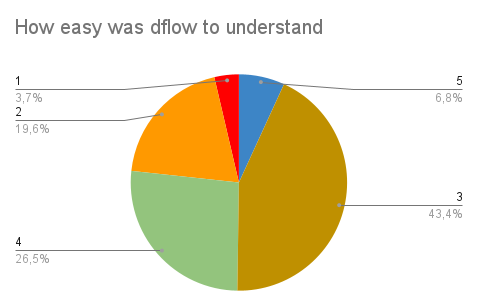}  
  \caption{Understanding of dFlow}
\end{subfigure}
\begin{subfigure}{.49\textwidth}
  \centering
  \includegraphics[width=\linewidth]{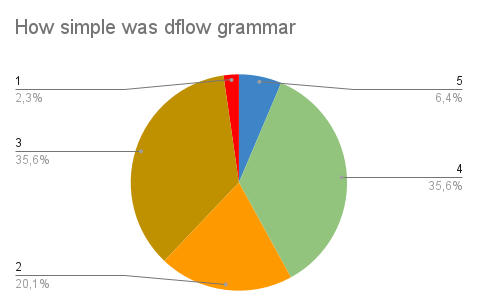}  
  \caption{Simplicity of dFlow grammar}
\end{subfigure}


\begin{subfigure}{.49\textwidth}
  \centering
  \includegraphics[width=\linewidth]{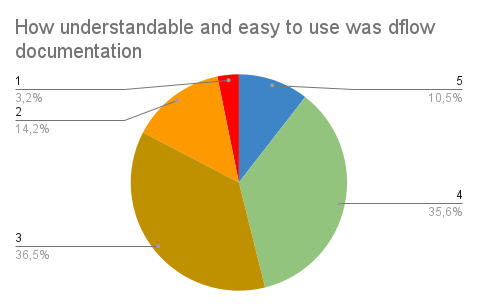}  
  \caption{Quality of dFlow documentation}
\end{subfigure}
\begin{subfigure}{.49\textwidth}
  \centering
  \includegraphics[width=\linewidth]{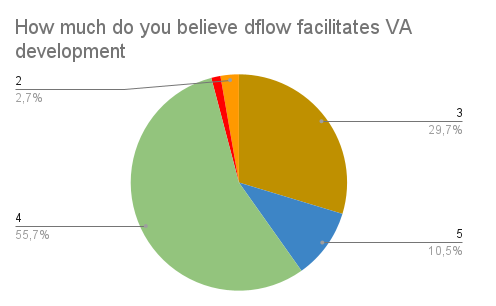}  
  \caption{How much does dFlow facilitate VA development}
\end{subfigure}


\begin{subfigure}{.49\textwidth}
  \centering
  \includegraphics[width=\linewidth]{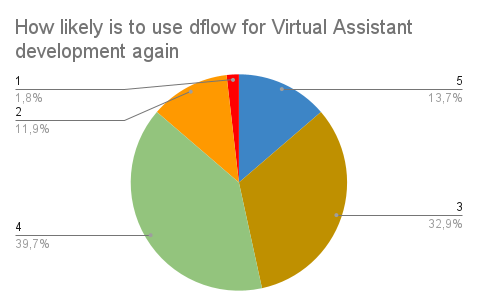}  
  \caption{How likely is to use dFlow for VA development again}
  \label{fig:reuse}
\end{subfigure}
\begin{subfigure}{.49\textwidth}
  \centering
  \includegraphics[width=\linewidth]{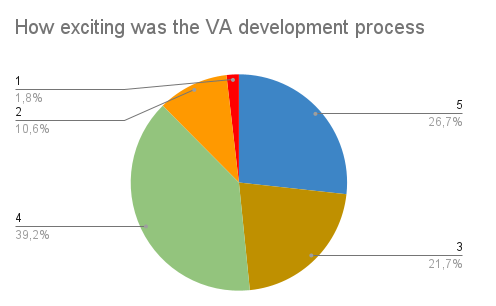}  
  \caption{How exciting was the VA development process}
  \label{fig:excitement}
\end{subfigure}
\caption{dFlow feedback} \label{dflow-feedback}
\end{figure}

\begin{figure}[!htb]
  \centering
  \includegraphics[width=\linewidth]{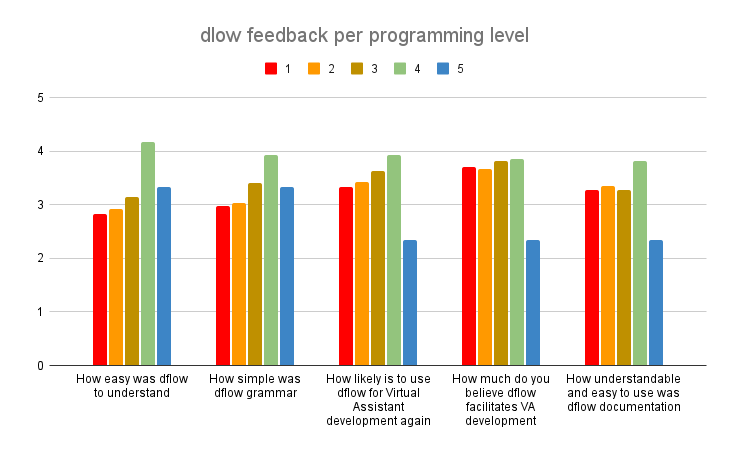}  
  \caption{dFlow feedback per programming level \label{dflow-feedback-per-level}}
\end{figure}

The overall understanding of dFlow is depicted in Figure \ref{dflow-feedback}. Most questions had a 1-5 grading score with 5 being the most positive answer. In all presented charts the following coloring format is used: red for 1, orange for 2, dark yellow for 3, green for 4, and blue for 5. We also grouped the participants into five clusters according to their stated programming level skillset and present the average answers for each of the five first questions in Figure \ref{dflow-feedback-per-level}. We see that more experienced participants (with a programming skills level of 3 and 4) reported a significantly better overall understanding of dFlow language and concepts than the rest of the workshop members, as anticipated. It is noted that there were only 3 participants who stated they had excellent programming knowledge (level 5) and one of those reported a less confident understanding of dFlow, hence the more mediocre average results in three out of the five questions. However, less skilled participants successfully comprehended the various concepts and they reported a slightly worse overall understanding than the mid-level participants but still adequate, even though they struggled in a few cases, which was expected since they significantly lack development experience. Consequently, we can argue that dFlow does accelerate the entire VA development process (R.Q.1) as less time is needed for domain knowledge comprehension and it does not restrict non-experienced from developing VAs (R.Q.2). In addition, the combination of high excitement for VA development (Figure \ref{fig:excitement}) with the high possibility of reusing dFlow (Figure \ref{fig:reuse}) can be due to the convenient dFlow interface (R.Q.3).

\begin{figure}[!htb]\captionsetup[subfigure]{font=footnotesize}
    
\begin{subfigure}{.49\textwidth}
  \centering
  \includegraphics[width=.9\linewidth]{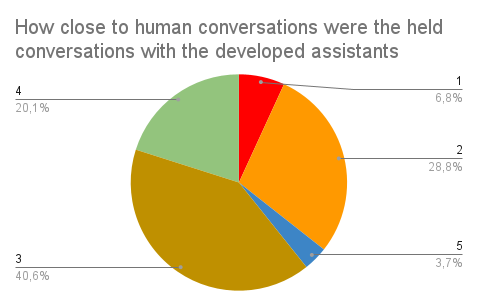}  
  \caption{How close to human conversations were the held conversations with the developed assistants}
\end{subfigure}
\begin{subfigure}{.49\textwidth}
  \centering
  \includegraphics[width=.9\linewidth]{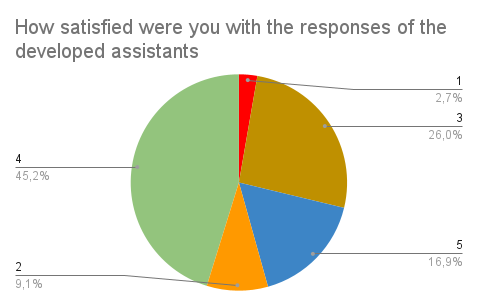}  
  \caption{How satisfied were you with the responses of the developed assistants}
\end{subfigure}


\begin{subfigure}{.48\textwidth}
  \centering
  \includegraphics[width=.9\linewidth]{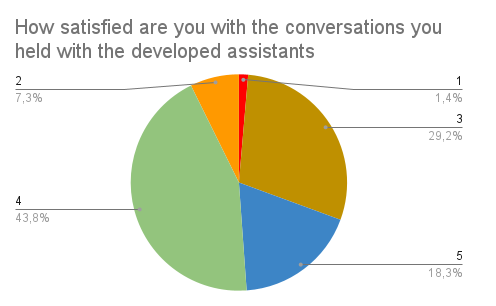}  
  \caption{How satisfied are you with the conversations you held with the developed assistants}
\end{subfigure}
\begin{subfigure}{.49\textwidth}
  \centering
  \includegraphics[width=.9\linewidth]{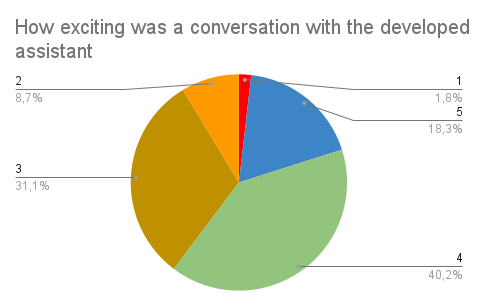}  
  \caption{How exciting was a conversation with the developed assistant}
\end{subfigure}
\caption{Developed assistants feedback \label{development-feedback}}
\end{figure}

Regarding the developed assistants, participants reported positive overall responses. Particularly, as presented in Figure \ref{development-feedback} the majority stated that the conversations held with the assistants were decent with respect to human conversations and their responses were adequate even though simple. This comment was anticipated since the deployed scenarios were rather simple and straightforward and only a small amount of trainable data were given. This choice was made as the aim of the workshop was to develop Minimum Viable Assistants rather than complex production-ready systems. Despite that, most participants were satisfied that they were able to chat with a model and became very interested in the field. 

Furthermore, the 219 participants designed 842 models throughout the workshop, 3 of which were deployed and tested by all participants. We grouped the models into the three use cases and measured the average and minimum time participants needed to develop each of their models, the average Lines of Code (LoC) written in dFlow, and the LoC of the generated Rasa models, and present the results in Table \ref{tab:dev-time}. We witness a huge reduction of time needed for creating and deploying VAs, as for example, the first and rather simple scenario required around 31 minutes to develop plus a few more for deployment, while the domain and concept knowledge was passed in the first part of the section. On the contrary, learning and using the Rasa framework would require around one full day of work, while the could-native setup would need at least one more. This reduction of development and deployment time from a few days to less than an hour results in a huge productivity gain and can facilitate dialogue-based application development. Similarly, an expert developer on both Rasa and dFlow was asked to implement the three scenarios with dFlow and Rasa and report the needed time which is presented in Table \ref{tab:master-dev-time}. It is clear that dFlow also benefits expert developers with respect to development time and effort, even in these simple scenarios.

\begin{table}[h]
\centering
\begin{tabular}{lllll} 
\toprule
\textbf{Scenario} & \textbf{Avg Time} & \textbf{Min Time} & \textbf{Avg LoC} & \textbf{Avg LoC} \\
& \textbf{Needed (min)} & \textbf{Needed (min)} & \textbf{in dFlow} & \textbf{in Rasa} \\ \midrule
\textit{Hello World} & 31 & 10 & 29 & 96 \\
\textit{User Profile} & 43 & 13 & 46 & 204 \\
\textit{Weather VA} & 40 & 13 & 63 & 253 \\
\bottomrule
\end{tabular}
\caption{Development Time (in minutes) and Lines of Code of Workshop Participants}
\label{tab:dev-time}
\end{table}

\begin{table}[h]
\centering
\begin{tabular}{lll} 
\toprule
\textbf{Scenario} & \textbf{Rasa Time} & \textbf{dFlow Time}\\ 
 & \textbf{Needed (min)} & \textbf{Needed (min)} \\
\midrule
\textit{Hello World} & 5 & 2 \\
\textit{User Profile} & 9 & 4 \\
\textit{Weather Assistant} & 13 & 5 \\
\bottomrule
\end{tabular}
\caption{Development Time (in minutes) of Rasa and dFlow Expert}
\label{tab:master-dev-time}
\end{table}

\subsection{Threats to validity}

The validity of our study, while thorough and extensive, has a few threats that should be discussed. First of all, one can argue that dFlow requires a decent level of programming skills and knowledge in order to comprehend the various dFlow concepts. It is true that a basic level of familiarity is needed and for this reason, dFlow aims to assist citizen developers, which do satisfy this requirement. According to the outcome of the study, their level of basic all-around knowledge was more than enough. Similarly, the proposed language could be considered complex and very difficult to understand and use. However, this is not supported by the findings as participants characterized it as simple and straightforward, while they succeeded in implementing several scenarios with minimum complexity struggles.

Furthermore, dFlow is based on the Rasa framework for generating the VAs, which currently is fully open-source. However, it is frequently updated with new features, which can leave dFlow far behind with respect to features and functionality, while it is always possible that one day it might become close-sourced. To that end, DFlow has been modeled in a hardware and framework-agnostic manner, as discussed in R.Q.2. Particularly, the dFlow meta-model is entirely independent of the used framework for generating the VA models, and hence, multiple generators can be implemented, one for each different framework or framework version. Consequently, as Rasa advances, a new dFlow generator can be developed that will include the new Rasa features, while in case Rasa becomes close-sourced, another framework can be employed for the generation and deployment of the VA models, with no modifications on the dFlow meta-model.

Finally, another threat may exist regarding the selection of textual representations for the DSL, instead of visual. One can argue that a graphical DSL could be better suited for citizen developers, in terms of rapid bootstrapping and productivity gain, while also syntax/grammar-related issues would be diminished. On the contrary, the outcome of our study showed that dFlow has a rather low barrier to understanding and using, while any issues faced were of small scale and did not prevent users from fully utilizing the language, aspects of textual DSLs that are also discussed in other studies \cite{xatkit}.

\section{Conclusions \& Future Work}
\label{sec:5}

In this work, we presented a textual DSL for rapid Virtual Assistant development called dFlow using open-source technologies. We demonstrated a cloud-native approach to further facilitate the development process and presented our system to 219 participants with small programming skills and minimum domain knowledge. The results showed that almost all participants were able to design and deploy VAs even though they were not familiar with NLP or VAs. We argue that this is expected to lead to a reduction in the development time and will allow for a wider penetration of Virtual Assistants in the market.

We are not hoping to replace the commonly used tools and frameworks, such as Rasa or Dialogflow, but to build on top of them and make them available in an integrated, vendor-agnostic, and system-agnostic context, opening the gates of conversational application production to citizen developers by reducing excessive hand-coding and domain-specific technical knowledge requirements. The proposed dFlow DSL covers the framework-specific and NLP-specific configuration details and focuses entirely on assistant creation that will be able to be integrated everywhere.

We aim to further improve user experience and build a complete end-to-end framework. First, we plan to expand the cloud-native approach by developing a cluster that can automate hosting and managing multiple Rasa instances. We are also discussing designing a visual DSL extending the existing one, as a different interface, and thus validating whether it is a more convenient design interface. Furthermore, we are planning on extending the supported functionalities of dFlow and include multi-turn dialogues, conditional logic, Machine Learning customization, and a complete error validator.

There are a few more possible steps that require extensive NLP and MDE research. An interesting feature is to automate the development of dFlow scenarios that use external services, such as OpenAPI. This can be accomplished by developing the OpenAPI-to-dFlow transformation, while the most difficult task could be to automate the intent example generation using a service description or metadata. In addition, an NLP-intensive task is the semantic merging of several dFlow models by employing state-of-the-art Language Models. Finally, similarly to another study \cite{perez19} a conversational interface with which we can describe dFlow models could be implemented. That would result in a further increase in user experience and could potentially enable citizen developers to easily design and deploy (seamless) Virtual Assistants.

\bibliographystyle{unsrt}  
\bibliography{main.bib}

\end{document}